\documentclass[%
aps,
prd,
floatfix,
preprintnumbers,
preprint,
showpacs,
nofootinbib,
tightenlines,
amssymb,
amsmath
]{revtex4-2}

\usepackage{ifpdf}
\ifpdf
	\usepackage{cmap} % search in PDF
	\usepackage[pdftex]{color,graphicx}% include graphics
	\usepackage{epstopdf} %eps. -> .pdf figures
\else
	\usepackage{color,graphicx}
\fi
\usepackage{bm}% bold math
\usepackage[dvipsnames]{xcolor}
\usepackage{hyperref}
\usepackage{longtable}
\hypersetup{%
colorlinks=true,%
citecolor=Blue,%
filecolor=Black,%
linkcolor=Blue,%
urlcolor=Violet
}
\ifpdf
\usepackage[activate={true,nocompatibility},final,tracking=true,kerning=true,spacing=true,factor=1100,stretch=10,shrink=10]{microtype}
\fi
\microtypecontext{spacing=nonfrench}

\newcommand{\gev}    {\:\mathrm{GeV}}
\newcommand{\mev}    {\:\mathrm{MeV}}
\newcommand{\gevsq}  {\:\mathrm{GeV}^2}

\newcommand{\ceps}{\varepsilon}

\newcommand{\average}[1]{\left\langle{#1}\right\rangle}
\newcommand{\la}{\left\langle}
\newcommand{\ra}{\right\rangle}
\newcommand{\eq}[1]{Eq.(\ref{#1})}

\begin{document}
\preprint{INR-TH-2026-003}
\title{%
On the Cancellation of Nuclear Effects in the Valence Region 
}

\author{S.~A.~Kulagin}
\email[]{kulagin.physics@gmail.com}
\affiliation{Institute for Nuclear Research of the Russian Academy of Sciences,
117312 Moscow, Russia}
\author{R.~Petti}
\email[]{roberto.petti@cern.ch}
\affiliation{Department of Physics and Astronomy, University of South Carolina, Columbia SC 29208, USA}

%%\date{\today}

\begin{abstract}
\noindent
Deep-inelastic $e/\mu$ scattering data on targets ranging from deuterium to lead indicate that nuclear modifications to the structure functions of bound nucleons are minimal in the kinematic region where valence quark distributions peak. A global analysis of measurements of the isoscalar cross-section ratios $\sigma^A/\sigma^{{}^2\text{H}}$ in the range $0.25 \leq x \leq 0.35$ reveals a remarkable cancellation of nuclear effects across all nuclei, yielding a mean value of $0.9988 \pm 0.0023$. We discuss these results and explore potential  interpretations within a microscopic model for the nuclear modifications of the structure functions.
\end{abstract}

%\begin{keyword}
%Parton distributions, nuclear parton distributions, Nuclear correction in deep-inelastic scattering, EMC effect
%\end{keyword}

%\pacs{13.60.Hb, 12.38.Qk} %!Fix pacs!

\maketitle
\newpage

\section{Introduction}
\label{sec:intro}

Deep-inelastic scattering (DIS) experiments on nuclear targets have established that the nuclear environment significantly modifies the bound-nucleon partonic structure, even at energy and momentum scales several orders of magnitude higher than those typical of nuclear ground-state processes (for reviews, see Refs.~\cite{Arneodo:1992wf,Norton:2003cb}). A precise knowledge of nuclear modifications to bound protons and neutrons is therefore essential for interpreting hard-scattering processes across a wide spectrum of experiments, including fixed-target DIS measurements at JLab~\cite{Dudek:2012vr}, studies at colliders like the LHC~\cite{Albacete:2017qng} and the future Electron-Ion Collider (EIC)~\cite{AbdulKhalek:2021gbh}, as well as measurements in long-baseline neutrino oscillation experiments such as DUNE~\cite{DUNE:2020ypp}.

\begin{figure}[htb!]
\centering
\includegraphics[width=0.8\textwidth]{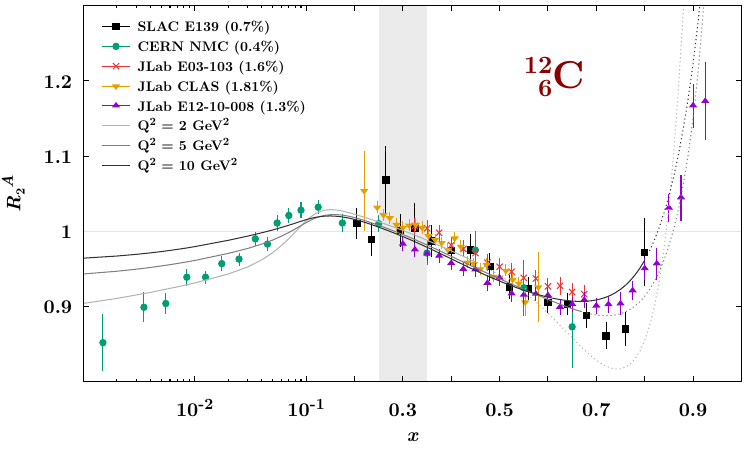}%
\caption{%
Illustration of the general behavior for the structure function ratios $R_2^A$. The data points represent the available measurements for the ${}^{12}$C nucleus from the CERN NMC~\cite{NewMuon:1996yuf}, SLAC E139~\cite{Gomez:1993ri}, JLab E03-103~\cite{Arrington:2021vuu}, JLab CLAS~\cite{CLAS:2019vsb}, and JLab E12-10-008~\cite{HallC:2022utd} experiments with a cut on the hadronic invariant mass $W \geq 1.8 \gev$. The error bars represent the quadrature sum of statistical and point-to-point systematic uncertainties. The normalization uncertainties are not included and are given in braces for completeness. The predictions from the microscopic nuclear model of Ref.~\cite{Kulagin:2004ie} are shown at $Q^2 = 2$, $5$, and $10\gevsq$ (solid lines for $W>1.8\gev$ and dotted lines for $W<1.8\gev$). The vertical shaded region highlights the kinematic range $0.25 \leq x \leq 0.35$ analyzed in the present work.
}  
\label{fig:emc-c12}
\end{figure}

Systematic investigations of nuclear effects in DIS have been conducted across diverse nuclear targets from deuterium to lead as a function of the Bjorken scaling variable $x$ and the four-momentum transfer squared $Q^2$~\cite{EuropeanMuon:1992pyr,BCDMS:1985dor,BCDMS:1987upi,Gomez:1993ri,Amaudruz:1995tq,NewMuon:1991exl,NewMuon:1996yuf,HERMES:1999bwb,HERMES:2002llm,Seely:2009gt,HallC:2022utd,CLAS:2019vsb,Griffioen:2015hxa,JeffersonLabHallATritium:2024las}. Experimental results are typically reported as structure-function ratios for a nucleus with mass number $A$ relative to the deuterium, $F_2^A(x,Q^2)/F_2^{{}^2\text{H}}(x,Q^2)$. Such ratios follow a universal pattern across the periodic table: a suppression at small $x < 0.05$ (nuclear shadowing), followed by a modest enhancement at $0.05 < x < 0.3$ (antishadowing), a nearly linear reduction in the valence quark region $0.3 < x < 0.65$ (EMC effect) reaching a local minimum around $x = 0.7\text{--}0.75$ (the dip region), and a steep rise at $x > 0.8$ driven by the nuclear momentum distributions~\cite{Arneodo:1992wf,Norton:2003cb}.
This characteristic behavior is illustrated in Fig.~\ref{fig:emc-c12} for the carbon nucleus. A combined logarithmic-linear scale is used to clearly display both the low- and high-$x$ features. Predictions from the microscopic model of Ref.~\cite{Kulagin:2004ie} are also shown for a few fixed $Q^2$ values. Notably, a pronounced $Q^2$-dependence appears in both the shadowing region and around the large-$x$ dip, whereas it is substantially suppressed throughout 
the EMC-effect region.%
\footnote{For an accurate analysis model predictions must be calculated for the exact kinematics of each data point, taking into account the $Q^2\text{--}x$ correlation in fixed-target experimental data~\cite{Kulagin:2004ie}. Note that in the nuclear shadowing region the data kinematics extends into the sub-GeV$^2$ regime ($Q^2 < 1\gevsq$).} 

A key observation emerging from global experimental data~\cite{Gomez:1993ri,Kulagin:2004ie,Kulagin:2010gd,Weinstein:2010rt} is the cancellation of nuclear corrections around $x \approx 0.3$, where valence quark distributions dominate. Minimal nuclear modification are observed within the interval $0.25 \leq x \leq 0.35$ across diverse targets and kinematic domains~\cite{EuropeanMuon:1992pyr,BCDMS:1985dor,BCDMS:1987upi,Gomez:1993ri,Amaudruz:1995tq,NewMuon:1991exl,NewMuon:1996yuf,HERMES:1999bwb,HERMES:2002llm,Seely:2009gt,HallC:2022utd,CLAS:2019vsb,Griffioen:2015hxa,JeffersonLabHallATritium:2024las}. A detailed study of this region offers valuable insight into the underlying interplay of distinct nuclear dynamics. Furthermore, characterizing this cancellation also provides a reliable reference point for the absolute normalization of existing and future measurements.

In this paper, we perform a quantitative assessment of the cancellation of nuclear effects using the world DIS data in the range $0.25 \leq x \leq 0.35$ (Sec.~\ref{sec:data}). We then compare these data with predictions from the microscopic nuclear model of Refs.~\cite{Kulagin:2004ie,Kulagin:2010gd,Kulagin:2014vsa} and discuss the physical mechanisms originating the cancellation of nuclear modifications within such a narrow kinematic window (Sec.~\ref{sec:discus}).

\section{Analysis of data}
\label{sec:data}

Experimental measurements of nuclear effects in deep-inelastic scattering (DIS) are typically reported as ratios of the cross-section on a nucleus with mass number $A$ to that on deuterium, $\sigma^A/\sigma^{{}^2\text{H}}$, normalized by the respective nucleon numbers. It is standard practice to assume that the cross-section ratio is equivalent to the ratio of structure functions, $R_2^A = F_2^A / F_2^{{}^2\text{H}}$, an assumption supported by experimental observations showing a similar behavior of the ratio $R = F_L/F_T$ of longitudinal to transverse structure functions in nuclei~\cite{Dasu:1993vk}. For isoscalar nuclei — those with an equal number of protons ($Z$) and neutrons ($N$) — the ratio $R_2^A$ would be unity if nuclear corrections were absent. Consequently, any deviation of $R_2^A$ from unity is interpreted as a manifestation of nuclear modifications to the bound nucleon structure functions.

For non-isoscalar nuclei ($Z \neq N$), the measured $F_2^A$ and $R_2^A$ are usually corrected for non-isoscalarity as follows:
{\allowdisplaybreaks
\begin{align}
\label{eq:f2is}
F_2^{A(\text{is})} &= F_2^A \frac{A F_2^N}{Z F_2^p + N F_2^n} = F_2^A \frac{A ( 1+F_2^n/F_2^p)}{2( Z + N F_2^n/F_2^p)},
\\
R_2^A &= F_2^{A(\text{is})} / F_2^{{}^2\text{H}},
\label{eq:r2}
\end{align}
}%
where $F_2^p$ and $F_2^n$ are the structure functions of the free proton and neutron, respectively, and $F_2^N = \frac{1}{2}(F_2^p + F_2^n)$. Throughout the following discussion, we also employ the ratio:
\begin{align}
\label{eq:ris}
R_2^{A(\text{is})} &= \frac{A F_2^A}{ZF_2^p + NF_2^n}=\frac{F_2^{A(\text{is})}} {F_2^N}.
\end{align}
While not directly measurable, the ratio $R_2^{A(\text{is})}$ provides a useful metric for quantifying the magnitude of nuclear effects in a generic nucleus.

Table~\ref{tab:avg} compiles the data on \(R_{2}^{A}\) in the kinematic region \(0.25\leq x \leq 0.35\). These data were obtained from multiple DIS experiments conducted at CERN, SLAC, DESY, and JLab over several decades~\cite{JeffersonLabHallATritium:2024las,Arrington:2021vuu,Seely:2009gt,HERMES:1999bwb,Gomez:1993ri,Amaudruz:1995tq,HallC:2022utd,NewMuon:1996yuf,CLAS:2019vsb,BCDMS:1985dor,Dasu:1993vk,BCDMS:1987upi,EuropeanMuon:1992pyr,Garutti:2003,Griffioen:2015hxa}. For each nuclear target and experiment \(k\), the data points in Table~\ref{tab:avg} are combined into the weighted average \(\overline{r}_k \pm \delta \overline{r}_k\):
\begin{align} 
\label{eq:avg1}
\overline{r}_k & = \frac{\sum_i w_i r_i}{\sum_i w_i}, 
\\ 
w_i & = \frac{1}{(\delta r_i^{\text{stat}})^2 + (\delta r_i^{\text{syst}})^2},
\\
\delta \overline{r}_k  & = \left[ \frac{1}{ \sum_i w_i} + (\delta r_k^{\text{norm}})^2 \right]^{1/2},
\end{align}
where $r$ represents the SF ratios listed in Table~\ref{tab:avg}, $\delta r_i^{\rm stat}$ is the statistical uncertainty, and $\delta r_i^{\rm syst}$ is the total point-to-point systematic uncertainty. The normalization uncertainty, $\delta r_k^{\rm norm}$, is treated as fully correlated across all points within a given experiment. For the majority of the experiments considered, $\delta r_k^{\rm norm}$ constitutes the dominant contribution to the total uncertainty $\delta \overline{r}_k$ (see Appendix~\ref{sec:datanorm}).%
\footnote{For an experiment in which the normalization uncertainty is larger than the combined statistical and point-to-point systematic uncertainties the number of data points considered has a marginal impact on the weighted average in \eq{eq:avg2}.}
When multiple experiments provide data for the same target, the individual averages were further combined into a final weighted average:
\begin{align} 
\label{eq:avg2} 
\overline{r} \pm \delta \overline{r}  & =  \frac{\sum_k w_k \overline{r}_k}{\sum_k w_k} \pm \frac{1}{\left( \sum_k w_k \right)^{1/2}},
\\
w_k & = 1/(\delta \overline{r}_k)^2 .
\end{align} 
The resulting averages for each target are listed in Table~\ref{tab:avg}. The recent $R_2^A$ data for $^3$H and $^3$He from the MARATHON experiment~\cite{JeffersonLabHallATritium:2024las} are included for completeness. However, they are excluded from the averages because the MARATHON analysis utilizes the observation that nuclear effects in light nuclei near $x \approx 0.3$ are minimal compared with the corresponding experimental uncertainties~\cite{JeffersonLabHallATritium:2024las}.

\begin{figure}[htb]
\centering
\includegraphics[width=1.00\textwidth]{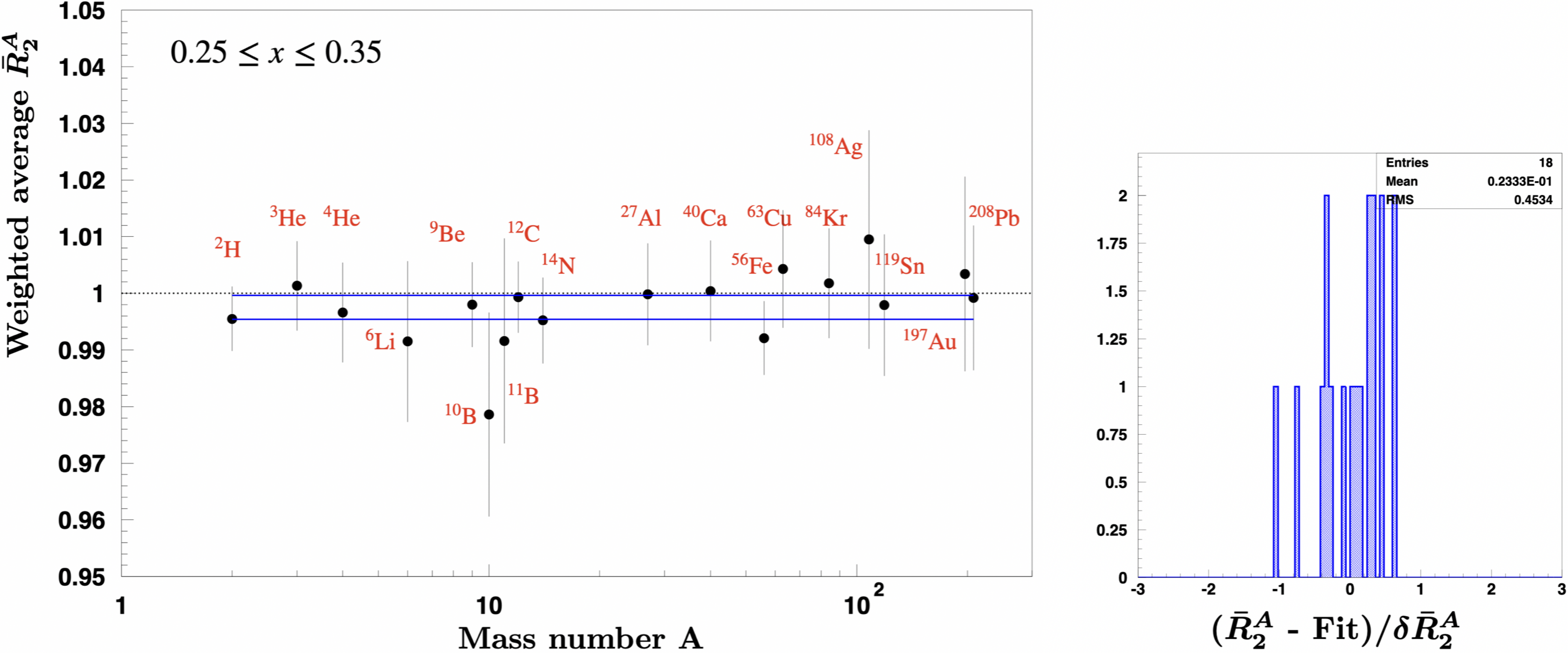}%
\caption{%
(Left) Weighted averages of $R_2^A$ measurements in the range $0.25 \leq x \leq 0.35$ (Table~\ref{tab:avg}). The measurements from the MARATHON experiment are not included. The solid lines represent the $\pm 1\sigma$ band obtained from a fit to all nuclei with a constant $R_0$ (Table~\ref{tab:fitsA}). (Right) Distribution of the deviations with respect to the fitted $R_0$ for the weighted averages of all 18 nuclei in the left panel, normalized to the corresponding experimental uncertainties. 
}  
\label{fig:emcx03}
\end{figure}

Figure~\ref{fig:emcx03} displays the combined averages as a function of the mass number $A$. To investigate potential $A$-dependence, we performed a fit using the functional form $R_0 + R_1 \log A$. The resulting slope, $R_1 = 0.0005 \pm 0.1549$, is consistent with zero, indicating a universal behavior of the structure function ratios across all nuclei in the range $0.25 \leq x \leq 0.35$. A fit with a constant value yields $R_0 = 0.9979 \pm 0.0023$ for the full data set (see Table~\ref{tab:fitsA}). All nuclei are in good agreement with the constant ratio $R_0$ within uncertainties, as demonstrated by the corresponding polls (Fig.~\ref{fig:emcx03}) and by a $\chi^2/\text{d.o.f}=0.21$. We also found remarkable consistency when fitting different subsets of nuclei, as summarized in Table~\ref{tab:fitsA}. For example, a fit to the light nuclei ${}^3$He and ${}^4$He yields $R_0 = 0.9992 \pm 0.0058$.

Non-isoscalarity corrections in published measurements frequently rely on different inputs for the structure-function ratio $R_{np} = F_2^n / F_2^p$ in \eq{eq:f2is}, introducing a potential source of model systematics. To quantify the impact of these differences, we recalculated the corrections using a consistent $R_{np}$ based on a parametrization of the MARATHON~\cite{JeffersonLabHallATritium:2021usd} and NMC~\cite{NewMuon:1991exl} measurements, evolved to the average $Q^2$ of each individual data point (Appendix~\ref{sec:F2n-F2p}).
These updated values are reported in the final column of Table~\ref{tab:avg}. A fit to these corrected data using a constant value  yields $R_0 = 0.9988 \pm 0.0023$, in excellent agreement with the results derived from the original published measurements. This stability is further reinforced by fits  restricted exclusively to purely isoscalar ($Z=N$) nuclei (Table~\ref{tab:fitsA}).

{\setlength{\tabcolsep}{10pt}
\begin{longtable}{l|c|c|c|c} 
\caption{%
Summary of experimental data for the isoscalar structure function ratios $R_2^A$ in the kinematic range $0.25 \leq x \leq 0.35$. The uncertainties of individual data points are calculated as the quadrature sum of statistical ($\delta r^{\rm stat}$), point-to-point systematic ($\delta r^{\rm syst}$), and normalization ($\delta r^{\rm norm}$) uncertainties. The last column provides results with consistent non-isoscalarity corrections applied to all data (see text). The weighted averages across different experiments are computed according to \eq{eq:avg2}, with normalization uncertainties treated as fully correlated within each experiment. The MARATHON data are shown for comparison but are excluded from the weighted averages.
}
\label{tab:avg}
\\
\hline
Measurement  & $\average{x}$  &  $\average{Q^2}$   &  $R_2^A$  & $R_2^A$ corr. \\
\hline
\endfirsthead
\hline
Measurement  & $\average{x}$  &  $\average{Q^2}$   &  $R_2^A$  & $R_2^A$ corr. \\
\hline
\endhead
\hline
\endfoot
\endlastfoot
${}^2$H/(p+n) JLab BONuS~\cite{Griffioen:2015hxa} & 0.273 & 1.39 & $0.997 \pm 0.008$  & $0.997 \pm 0.008$ \\ 
${}^2$H/(p+n) JLab BONuS~\cite{Griffioen:2015hxa} & 0.323 & 1.50 & $0.994 \pm 0.008$  &  $0.994 \pm 0.008$ \\ 
\hline
\multicolumn{3}{l|}{Average of ${}^2$H/(p+n) measurements}  &  $0.9955 \pm 0.0057$  &  $0.9955 \pm 0.0057$ \\ % 
 \hline
 ${}^3$H/${}^2$H JLab MARATHON~\cite{JeffersonLabHallATritium:2024las}    & 0.285  &  3.99  &   $1.003 \pm 0.010$  &    $1.003 \pm 0.010$    \\ 
 ${}^3$H/${}^2$H JLab MARATHON~\cite{JeffersonLabHallATritium:2024las}    & 0.315   &  4.41  &   $0.992 \pm 0.010$  &   $0.992 \pm 0.010$   \\ 
 ${}^3$H/${}^2$H JLab MARATHON~\cite{JeffersonLabHallATritium:2024las}    & 0.345   &   4.83  & $0.999 \pm 0.011$  &   $0.999 \pm 0.011$  \\ 
 \hline
\multicolumn{3}{l|}{Average of ${}^3$H/${}^2$H measurements}  &  $0.9980 \pm 0.0073$  &  $0.9980 \pm 0.0073$ \\ % 
 \hline
${}^3$He/${}^2$H JLab E03-103~\cite{Arrington:2021vuu,Seely:2009gt}  &  0.325  &    2.87  &  $0.979 \pm 0.022$  &    $ 0.971 \pm 0.022$ \\ 
${}^3$He/${}^2$H JLab E03-103~\cite{Arrington:2021vuu,Seely:2009gt}  &  0.350  &    3.05  &  $0.975 \pm 0.022$  &    $ 0.967 \pm 0.022$  \\ 
${}^3$He/${}^2$H DESY HERMES~\cite{HERMES:1999bwb} & 0.250 & 3.07 & $1.008 \pm 0.012$ &   $1.008 \pm 0.012$ \\ 
${}^3$He/${}^2$H DESY HERMES~\cite{HERMES:1999bwb} & 0.350 & 4.13 & $1.004 \pm 0.011$ &   $1.004 \pm 0.011$ \\ 
${}^3$He/${}^2$H JLab MARATHON~\cite{JeffersonLabHallATritium:2024las}    & 0.285  &  3.99  &   $1.003 \pm 0.010$  &    $1.003 \pm 0.010$    \\ 
${}^3$He/${}^2$H JLab MARATHON~\cite{JeffersonLabHallATritium:2024las}    & 0.315   &  4.41  &   $0.992 \pm 0.010$  &   $0.992 \pm 0.010$   \\ 
${}^3$He/${}^2$H JLab MARATHON~\cite{JeffersonLabHallATritium:2024las}    & 0.345   &   4.83  & $0.999 \pm 0.011$  &   $0.999 \pm 0.011$  \\ 
\hline
 \multicolumn{3}{l|}{Average of ${}^3$He/${}^2$H measurements}  &  $1.0013 \pm 0.0079$  & $1.0001 \pm 0.0079$  \\  % 
 \hline
${}^4$He/${}^2$H JLab E03-103~\cite{Arrington:2021vuu,Seely:2009gt}  &  0.325  &    2.87  &   $ 1.005 \pm 0.018$   &   $1.005 \pm 0.018$ \\ 
${}^4$He/${}^2$H JLab E03-103~\cite{Arrington:2021vuu,Seely:2009gt}  &  0.350  &    3.05  &  0.995 $ \pm 0.018$   &   $0.995 \pm 0.018$  \\ 
${}^4$He/${}^2$H SLAC E139~\cite{Gomez:1993ri} &  0.265  &  5.00  &  $1.031 \pm 0.039$    &   $1.031 \pm 0.039$  \\ 
${}^4$He/${}^2$H SLAC E139~\cite{Gomez:1993ri} &  0.295  &  5.00  &  $1.004 \pm 0.031$    &   $1.004 \pm 0.031$ \\  
${}^4$He/${}^2$H SLAC E139~\cite{Gomez:1993ri} &  0.325  &  5.00  &  $ 0.991 \pm 0.035$    &    $ 0.991 \pm 0.035$ \\  
${}^4$He/${}^2$H CERN NMC~\cite{Amaudruz:1995tq} &  0.250 &  19.00 &  $0.996 \pm 0.013$   &    $0.996 \pm 0.013$ \\ 
${}^4$He/${}^2$H CERN NMC~\cite{Amaudruz:1995tq} &  0.350 &  24.00 &  $0.987 \pm 0.020$   &    $0.987 \pm 0.020$ \\  
\hline 
\multicolumn{3}{l|}{Average of ${}^4$He/${}^2$H measurements} &  $0.9966 \pm 0.0088$  &   $0.9966 \pm 0.0088$ \\  % 
\hline
${}^6$Li/${}^2$H CERN NMC~\cite{Amaudruz:1995tq}  &  0.250 &  17.00 &  $0.992 \pm 0.016$   &   $0.992 \pm 0.016$  \\ 
${}^6$Li/${}^2$H CERN NMC~\cite{Amaudruz:1995tq}  &  0.350 &  23.00 &  $0.990 \pm 0.027$   &   $0.990 \pm 0.027$ \\  
\hline 
\multicolumn{3}{l|}{Average of ${}^6$Li/${}^2$H measurements}  &  $0.9915 \pm 0.0142$  & $0.9915 \pm 0.0142$  \\   % 
\hline
${}^9$Be/${}^2$H JLab E03-103~\cite{Arrington:2021vuu,Seely:2009gt}  &  0.325  &    2.87  &  $1.021 \pm 0.021$  &   $ 1.025 \pm 0.021$  \\ 
${}^9$Be/${}^2$H JLab E03-103~\cite{Arrington:2021vuu,Seely:2009gt}  &  0.350  &    3.05  &  $1.013 \pm 0.021$  &   $ 1.016 \pm 0.021$ \\ 
${}^9$Be/${}^2$H JLab E12-10-008~\cite{HallC:2022utd} & 0.300  &  4.26  &   $0.985 \pm 0.019$  &  $0.988 \pm 0.019$  \\   
${}^9$Be/${}^2$H JLab E12-10-008~\cite{HallC:2022utd} & 0.325  &  4.51  &   $0.972 \pm 0.019$  &   $0.975 \pm 0.019$  \\    
${}^9$Be/${}^2$H SLAC E139~\cite{Gomez:1993ri} &  0.265  &  5.00  &  $1.012 \pm 0.028$    &  $1.021 \pm 0.028$  \\ 
${}^9$Be/${}^2$H SLAC E139~\cite{Gomez:1993ri} &  0.295  &  5.00  &  $1.003 \pm 0.018$    &   $1.012 \pm 0.018$ \\  
${}^9$Be/${}^2$H SLAC E139~\cite{Gomez:1993ri} &  0.325  &  5.00  &  $ 0.992 \pm 0.023$    &  $1.001 \pm 0.023$  \\  
${}^9$Be/${}^{12}$C$\times$${}^{12}$C/${}^2$H NMC~\cite{Amaudruz:1995tq,NewMuon:1996yuf} &  0.250 &  35.50 &  $1.011 \pm 0.018$   &  $1.011 \pm 0.018$  \\ 
${}^9$Be/${}^{12}$C$\times$${}^{12}$C/${}^2$H NMC~\cite{Amaudruz:1995tq,NewMuon:1996yuf} &  0.350 &  45.40 &  $0.972 \pm 0.024$   &   $0.972 \pm 0.024$  \\  
\hline
 \multicolumn{3}{l|}{Average of ${}^9$Be/${}^2$H measurements}  &  $0.9980 \pm 0.0075$ &  $ 1.0025 \pm 0.0075$  \\   % 
 \hline
${}^{10}$B/${}^2$H JLab E12-10-008~\cite{HallC:2022utd} & 0.300  &  4.26  &   $0.982 \pm 0.019$  &   $0.982 \pm 0.019$  \\   
${}^{10}$B/${}^2$H JLab E12-10-008~\cite{HallC:2022utd} & 0.325  &  4.51  &   $0.975 \pm 0.019$  &   $0.975 \pm 0.019$  \\    
\hline
\multicolumn{3}{l|}{Average of ${}^{10}$B/${}^2$H measurements}  &  $0.9786 \pm 0.0180$ &  $0.9785 \pm 0.0180$  \\  % 
 \hline
${}^{11}$B/${}^2$H JLab E12-10-008~\cite{HallC:2022utd} & 0.300  &  4.26  &   $0.993 \pm 0.019$  &  $0.995 \pm 0.019$  \\   
${}^{11}$B/${}^2$H JLab E12-10-008~\cite{HallC:2022utd} & 0.325  &  4.51  &   $0.990 \pm 0.019$  &  $0.992 \pm 0.019$  \\    
\hline
\multicolumn{3}{l|}{Average of ${}^{11}$B/${}^2$H measurements}  &  $0.9916 \pm 0.0181$  &  $0.9936 \pm 0.0181$ \\   % 
 \hline
${}^{12}$C/${}^2$H JLab E03-103~\cite{Arrington:2021vuu,Seely:2009gt}  &  0.325  &    2.87  &  $1.007 \pm 0.020$  &   $1.007 \pm 0.020$ \\ 
${}^{12}$C/${}^2$H JLab E03-103~\cite{Arrington:2021vuu,Seely:2009gt}  &  0.350  &    3.05  &  $1.004 \pm 0.020$  &   $1.004 \pm 0.020$ \\ 
${}^{12}$C/${}^2$H JLab E12-10-008~\cite{HallC:2022utd} & 0.300  &  4.26  &   $0.983 \pm 0.019$  &   $0.983 \pm 0.019$ \\   
${}^{12}$C/${}^2$H JLab E12-10-008~\cite{HallC:2022utd} & 0.325  &  4.51  &   $0.975 \pm 0.019$  &   $0.975 \pm 0.019$ \\    
${}^{12}$C/${}^2$H JLab CLAS~\cite{CLAS:2019vsb} & 0.273  &  1.81   &  $1.018 \pm 0.020$  &   $1.018 \pm 0.020$ \\
${}^{12}$C/${}^2$H JLab CLAS~\cite{CLAS:2019vsb} & 0.287  &  1.86   &  $1.009 \pm 0.020$  &  $1.009 \pm 0.020$ \\
${}^{12}$C/${}^2$H JLab CLAS~\cite{CLAS:2019vsb} & 0.300  &  1.90   &  $1.005 \pm 0.020$  &   $1.005 \pm 0.020$ \\
${}^{12}$C/${}^2$H JLab CLAS~\cite{CLAS:2019vsb} & 0.313  &  1.94   &  $1.008 \pm 0.020$  &   $1.008 \pm 0.020$ \\
${}^{12}$C/${}^2$H JLab CLAS~\cite{CLAS:2019vsb} & 0.327  &  1.98   &  $1.009 \pm 0.020$  &   $1.009 \pm 0.020$ \\
${}^{12}$C/${}^2$H SLAC E139~\cite{Gomez:1993ri} &  0.265  &  5.00  &  $1.068 \pm 0.045$    &  $1.068 \pm 0.045$ \\ 
${}^{12}$C/${}^2$H SLAC E139~\cite{Gomez:1993ri} &  0.295  &  5.00  &  $1.001 \pm 0.022$    &  $1.001 \pm 0.022$  \\  
${}^{12}$C/${}^2$H SLAC E139~\cite{Gomez:1993ri} &  0.325  &  5.00  &  $1.004 \pm 0.033$    &   $1.004 \pm 0.033$  \\  
${}^{12}$C/${}^2$H CERN NMC~\cite{Amaudruz:1995tq} &  0.250 &  20.00 &  $1.010 \pm 0.011$   &   $1.010 \pm 0.011$ \\ 
${}^{12}$C/${}^2$H CERN NMC~\cite{Amaudruz:1995tq} &  0.350 &  27.00 &  $0.971 \pm 0.016$   &   $0.971 \pm 0.016$ \\  
\hline
\multicolumn{3}{l|}{Average of ${}^{12}$C/${}^2$H measurements}  &  $0.9993 \pm 0.0063$  &  $0.9993 \pm 0.0063$  \\  % 
 \hline
${}^{14}$N/${}^2$H DESY HERMES~\cite{HERMES:1999bwb} & 0.250 & 3.07 & $1.010 \pm 0.011$ &   $1.010 \pm 0.011$ \\ 
${}^{14}$N/${}^2$H DESY HERMES~\cite{HERMES:1999bwb} & 0.350 & 4.13 & $0.978 \pm 0.011$ &   $0.978 \pm 0.011$ \\ 
${}^{14}$N/${}^2$H CERN BCDMS~\cite{BCDMS:1985dor} &  0.275  &   64.5    &  $1.024 \pm 0.025$  &   $1.024 \pm 0.025$ \\
${}^{14}$N/${}^2$H CERN BCDMS~\cite{BCDMS:1985dor} &  0.350  &   86.5    &  $0.983 \pm 0.024$  &   $0.983 \pm 0.024$ \\
\hline
 \multicolumn{3}{l|}{Average of ${}^{14}$N/${}^2$H measurements}  &  $0.9952 \pm 0.0076$ &  $0.9952 \pm 0.0076$  \\   % 
 \hline
${}^{27}$Al/${}^2$H JLab CLAS~\cite{CLAS:2019vsb} & 0.273  &  1.81   &  $1.005 \pm 0.020$  &   $1.005 \pm 0.020$ \\
${}^{27}$Al/${}^2$H JLab CLAS~\cite{CLAS:2019vsb} & 0.287  &  1.86   &  $1.004 \pm 0.020$  &   $1.004 \pm 0.020$ \\
${}^{27}$Al/${}^2$H JLab CLAS~\cite{CLAS:2019vsb} & 0.300  &  1.90   &  $1.001 \pm 0.020$  &   $1.001 \pm 0.020$ \\
${}^{27}$Al/${}^2$H JLab CLAS~\cite{CLAS:2019vsb} & 0.313  &  1.94   &  $0.997 \pm 0.020$  &   $0.997 \pm 0.020$ \\
${}^{27}$Al/${}^2$H JLab CLAS~\cite{CLAS:2019vsb} & 0.327  &  1.98   &  $1.002 \pm 0.020$  &   $1.002 \pm 0.020$ \\
${}^{27}$Al/${}^2$H SLAC E139~\cite{Gomez:1993ri} &  0.265  &  5.00  &  $1.019 \pm 0.028$    &  $1.022 \pm 0.028$ \\ 
${}^{27}$Al/${}^2$H SLAC E139~\cite{Gomez:1993ri} &  0.295  &  5.00  &  $1.009 \pm 0.018$    &  $1.012 \pm 0.018$ \\  
${}^{27}$Al/${}^2$H SLAC E139~\cite{Gomez:1993ri} &  0.325  &  5.00  &  $0.975 \pm 0.023$    &  $0.978 \pm 0.023$  \\  
${}^{27}$Al/${}^{12}$C$\times$${}^{12}$C/${}^2$H NMC~\cite{Amaudruz:1995tq,NewMuon:1996yuf} &  0.250 &  35.50 &  $1.009 \pm 0.023$   &  $1.009 \pm 0.023$  \\ 
${}^{27}$Al/${}^{12}$C$\times$${}^{12}$C/${}^2$H NMC~\cite{Amaudruz:1995tq,NewMuon:1996yuf} &  0.350 &  45.30 &  $0.954 \pm 0.034$   &  $0.954 \pm 0.034$  \\  
\hline 
\multicolumn{3}{l|}{Average of ${}^{27}$Al/${}^2$H measurements}  &  $0.9998 \pm 0.0090$  &  $1.0014 \pm 0.0090$ \\   % 
\hline
${}^{40}$Ca/${}^2$H SLAC E139~\cite{Gomez:1993ri} &  0.265  &  5.00  &  $0.995 \pm 0.043$    &  $0.995 \pm 0.043$ \\ 
${}^{40}$Ca/${}^2$H SLAC E139~\cite{Gomez:1993ri} &  0.295  &  5.00  &  $1.037 \pm 0.024$    &  $1.037 \pm 0.024$ \\  
${}^{40}$Ca/${}^2$H SLAC E139~\cite{Gomez:1993ri} &  0.325  &  5.00  &  $0.996 \pm 0.034$    &   $0.996 \pm 0.034$ \\  
${}^{40}$Ca/${}^2$H CERN NMC~\cite{Amaudruz:1995tq} &  0.250 &  19.00 &  $0.993 \pm 0.012$   &  $0.993 \pm 0.012$ \\ 
${}^{40}$Ca/${}^2$H CERN NMC~\cite{Amaudruz:1995tq} &  0.350 &  24.00 &  $0.993 \pm 0.018$   &   $0.993 \pm 0.018$ \\  
\hline 
\multicolumn{3}{l|}{Average of ${}^{40}$Ca/${}^2$H measurements}  &  $1.0004 \pm 0.0089$  &  $1.0004 \pm 0.0089$ \\   % 
\hline
${}^{56}$Fe/${}^2$H JLab CLAS~\cite{CLAS:2019vsb} & 0.273  &  1.81   &  $1.017 \pm 0.020$  &   $1.017 \pm 0.020$ \\
${}^{56}$Fe/${}^2$H JLab CLAS~\cite{CLAS:2019vsb} & 0.287  &  1.86   &  $1.010 \pm 0.020$  &   $1.010 \pm 0.020$ \\
${}^{56}$Fe/${}^2$H JLab CLAS~\cite{CLAS:2019vsb} & 0.300  &  1.90   &  $1.005 \pm 0.020$  &   $1.005 \pm 0.020$ \\
${}^{56}$Fe/${}^2$H JLab CLAS~\cite{CLAS:2019vsb} & 0.313  &  1.94   &  $1.006 \pm 0.020$  &   $1.006 \pm 0.020$ \\
${}^{56}$Fe/${}^2$H JLab CLAS~\cite{CLAS:2019vsb} & 0.327  &  1.98   &  $1.006 \pm 0.020$  &   $1.006 \pm 0.020$ \\
${}^{56}$Fe/${}^2$H SLAC E139~\cite{Gomez:1993ri} &  0.265  &  5.00  &  $0.990 \pm 0.025$    &  $0.996 \pm 0.025$ \\ 
${}^{56}$Fe/${}^2$H SLAC E139~\cite{Gomez:1993ri} &  0.295  &  5.00  &  $0.992 \pm 0.017$    &  $0.998 \pm 0.017$ \\  
${}^{56}$Fe/${}^2$H SLAC E139~\cite{Gomez:1993ri} &  0.325  &  5.00  &  $0.981 \pm 0.022$    &   $0.987 \pm 0.022$ \\  
${}^{56}$Fe/${}^2$H SLAC E140~\cite{Dasu:1993vk} &  0.350  &  1.50  &  $1.000 \pm 0.012$    &  $1.004 \pm 0.012$ \\ 
${}^{56}$Fe/${}^2$H SLAC E140~\cite{Dasu:1993vk} &  0.350  &  2.50  &  $0.993 \pm 0.012$    &  $0.998 \pm 0.012$  \\  
${}^{56}$Fe/${}^2$H SLAC E140~\cite{Dasu:1993vk} &  0.350  &  5.00  &  $0.980 \pm 0.012$    &  $0.986 \pm 0.012$  \\ 
${}^{56}$Fe/${}^{12}$C$\times$${}^{12}$C/${}^2$H NMC~\cite{Amaudruz:1995tq,NewMuon:1996yuf} &  0.250 &  35.70 &  $1.011 \pm 0.018$   &  $1.011 \pm 0.018$  \\ 
${}^{56}$Fe/${}^{12}$C$\times$${}^{12}$C/${}^2$H NMC~\cite{Amaudruz:1995tq,NewMuon:1996yuf} &  0.350 &  45.40 &  $0.961 \pm 0.026$   &  $0.961 \pm 0.026$ \\  
${}^{56}$Fe/${}^2$H CERN BCDMS~\cite{BCDMS:1987upi} &  0.275  &   64.5    &  $1.000 \pm 0.021$  &  $1.007 \pm 0.021$  \\
${}^{56}$Fe/${}^2$H CERN BCDMS~\cite{BCDMS:1987upi} &  0.350  &   86.5    &  $0.959 \pm 0.021$  &  $0.967 \pm 0.021$ \\
\hline 
\multicolumn{3}{l|}{Average of ${}^{56}$Fe/${}^2$H measurements}  &  $0.9921 \pm 0.0065$  & $0.9960 \pm 0.0065$  \\  % 
\hline
${}^{63}$Cu/${}^2$H JLab E03-103~\cite{Arrington:2021vuu}  &  0.325  &    2.87  &  $1.028 \pm 0.023$  &    $1.032 \pm 0.023$ \\ 
${}^{63}$Cu/${}^2$H JLab E03-103~\cite{Arrington:2021vuu}  &  0.350  &    3.05  &  $1.014 \pm 0.023$  &    $ 1.017 \pm 0.023$  \\ 
${}^{63}$Cu/${}^2$H CERN EMC~\cite{EuropeanMuon:1992pyr} &  0.243 &  19.30 &  $1.023 \pm 0.015$   &  $1.023 \pm 0.015$ \\ 
${}^{63}$Cu/${}^2$H CERN EMC~\cite{EuropeanMuon:1992pyr} &  0.343 &  25.80 &  $0.956 \pm 0.020$   &  $0.956 \pm 0.020$ \\ 
\hline 
\multicolumn{3}{l|}{Average of ${}^{63}$Cu/${}^2$H measurements}  &  $1.0043 \pm 0.0104$  & $ 1.0052 \pm 0.0104$  \\  % 
\hline
${}^{84}$Kr/${}^2$H DESY HERMES~\cite{Garutti:2003} & 0.250 & 3.10 & $1.014 \pm 0.011$ &  $1.014 \pm 0.011$ \\ 
${}^{84}$Kr/${}^2$H DESY HERMES~\cite{Garutti:2003} & 0.350 & 4.14 & $0.966 \pm 0.018$ &  $0.966 \pm 0.018$  \\ 
\hline
 \multicolumn{3}{l|}{Average of ${}^{84}$Kr/${}^2$H measurements}  &  $1.0018 \pm 0.0097$  &  $1.0018 \pm 0.0097$ \\  % 
 \hline
${}^{108}$Ag/${}^2$H SLAC E139~\cite{Gomez:1993ri} &  0.265  &  5.00  &  $0.938 \pm 0.044$    & $0.9480 \pm 0.044$  \\ 
${}^{108}$Ag/${}^2$H SLAC E139~\cite{Gomez:1993ri} &  0.295  &  5.00  &  $1.030 \pm 0.025$    & $1.041 \pm 0.025$ \\  
${}^{108}$Ag/${}^2$H SLAC E139~\cite{Gomez:1993ri} &  0.325  &  5.00  &  $0.998 \pm 0.036$    &  $1.009 \pm 0.036$ \\  
\hline 
\multicolumn{3}{l|}{Average of ${}^{108}$Ag/${}^2$H measurements}  &  $1.0095 \pm 0.0193$  &  $1.0204 \pm 0.0193$ \\   % 
\hline
${}^{119}$Sn/${}^{12}$C$\times$${}^{12}$C/${}^2$H NMC~\cite{Amaudruz:1995tq,NewMuon:1996yuf} &  0.250 &  20.80 &  $1.015 \pm 0.015$   &  $1.015 \pm 0.015$  \\ 
${}^{119}$Sn/${}^{12}$C$\times$${}^{12}$C/${}^2$H NMC~\cite{Amaudruz:1995tq,NewMuon:1996yuf} &  0.350 &  26.60 &  $0.963 \pm 0.021$   &  $0.963 \pm 0.021$  \\  
\hline 
\multicolumn{3}{l|}{Average of ${}^{119}$Sn/${}^2$H measurements}  &  $0.9979 \pm 0.0125$  & $0.9979 \pm 0.0125$  \\  % 
\hline
${}^{197}$Au/${}^2$H JLab E03-103~\cite{Arrington:2021vuu}  &  0.325  &    2.87  &  $1.025 \pm 0.026$  & 1.030   $ \pm 0.026$ \\ 
${}^{197}$Au/${}^2$H JLab E03-103~\cite{Arrington:2021vuu}  &  0.350  &    3.05  &  $1.005 \pm 0.025$  &    $ 1.010 \pm 0.025$  \\ 
${}^{197}$Au/${}^2$H SLAC E139~\cite{Gomez:1993ri} &  0.265  &  5.00  &  $0.984 \pm 0.038$    &  $1.000 \pm 0.038$  \\ 
${}^{197}$Au/${}^2$H SLAC E139~\cite{Gomez:1993ri} &  0.295  &  5.00  &  $0.993 \pm 0.029$    &  $1.010 \pm 0.029$ \\  
${}^{197}$Au/${}^2$H SLAC E139~\cite{Gomez:1993ri} &  0.325  &  5.00  &  $0.984 \pm 0.035$    &   $1.001 \pm 0.035$  \\  
\hline 
\multicolumn{3}{l|}{Average of ${}^{197}$Au/${}^2$H measurements}  &  $1.0034 \pm 0.0172$  & 1.0061 $\pm 0.0172$ \\  % 
\hline
${}^{208}$Pb/${}^2$H JLab CLAS~\cite{CLAS:2019vsb} & 0.273  &  1.81   &  $1.024 \pm 0.021$  &   $1.024 \pm 0.021$ \\
${}^{208}$Pb/${}^2$H JLab CLAS~\cite{CLAS:2019vsb} & 0.287  &  1.86   &  $1.019 \pm 0.021$  &   $1.019 \pm 0.021$ \\
${}^{208}$Pb/${}^2$H JLab CLAS~\cite{CLAS:2019vsb} & 0.300  &  1.90   &  $1.012 \pm 0.021$  &   $1.012 \pm 0.021$  \\
${}^{208}$Pb/${}^2$H JLab CLAS~\cite{CLAS:2019vsb} & 0.313  &  1.94   &  $1.010 \pm 0.021$  &   $1.010 \pm 0.021$ \\
${}^{208}$Pb/${}^2$H JLab CLAS~\cite{CLAS:2019vsb} & 0.327  &  1.98   &  $1.011 \pm 0.021$  &   $1.011 \pm 0.021$ \\
${}^{208}$Pb/${}^{12}$C$\times$${}^{12}$C/${}^2$H NMC~\cite{Amaudruz:1995tq,NewMuon:1996yuf}~ &  0.250 &  20.80 &  $1.006 \pm 0.020$   &  $1.006 \pm 0.020$  \\ 
${}^{208}$Pb/${}^{12}$C$\times$${}^{12}$C/${}^2$H NMC~\cite{Amaudruz:1995tq,NewMuon:1996yuf} &  0.350 &  26.60 &  $0.945 \pm 0.030$   &  $0.945 \pm 0.030$  \\  
\hline 
\multicolumn{3}{l|}{Average of ${}^{208}$Pb/${}^2$H measurements}  &  $0.9992 \pm 0.0128$  &  $0.9992 \pm 0.0128$  \\  % 
\hline
\end{longtable}
}
\vspace*{-2ex} 

\begin{table}[!ht]
\caption{%
Results of a fit to the weighted averages of $R_2^A$ and $R_2^A$ corr. with a constant $R_0$ in the range $0.25 \leq x \leq 0.35$ for different subsets of nuclei. The measurements from the MARATHON experiment are not included in the weighted averages.
}
\label{tab:fitsA}
\begin{center}
{\setlength{\tabcolsep}{10pt}
\begin{tabular}{l|c} \hline
Fitted data  & $R_0$     \\
\hline
\multicolumn{2}{c}{$R_2^A$} \\ 
All nuclei                      & $0.9975 \pm 0.0021$              \\  
All nuclei except ${}^2$H       & $0.9979 \pm 0.0023$ \\ 
Only isoscalar nuclei           & $0.9964 \pm 0.0030$              \\ 
Only ${}^3$He, ${}^4$He & $0.9992 \pm 0.0058$            \\ 
\hline
\multicolumn{2}{c}{$R_2^A$ corr.\ } \\ 
All nuclei                      & $0.9983 \pm 0.0021$               \\  
All nuclei except ${}^2$H       & $0.9988 \pm 0.0023$ \\ 
Only ${}^3$He, ${}^4$He & $0.9985 \pm 0.0058$             \\
\hline
\end{tabular}
}
\vspace*{-5ex} 
\end{center}
\end{table}

\begin{figure}[htb]
\centering
\includegraphics[width=1.00\textwidth]{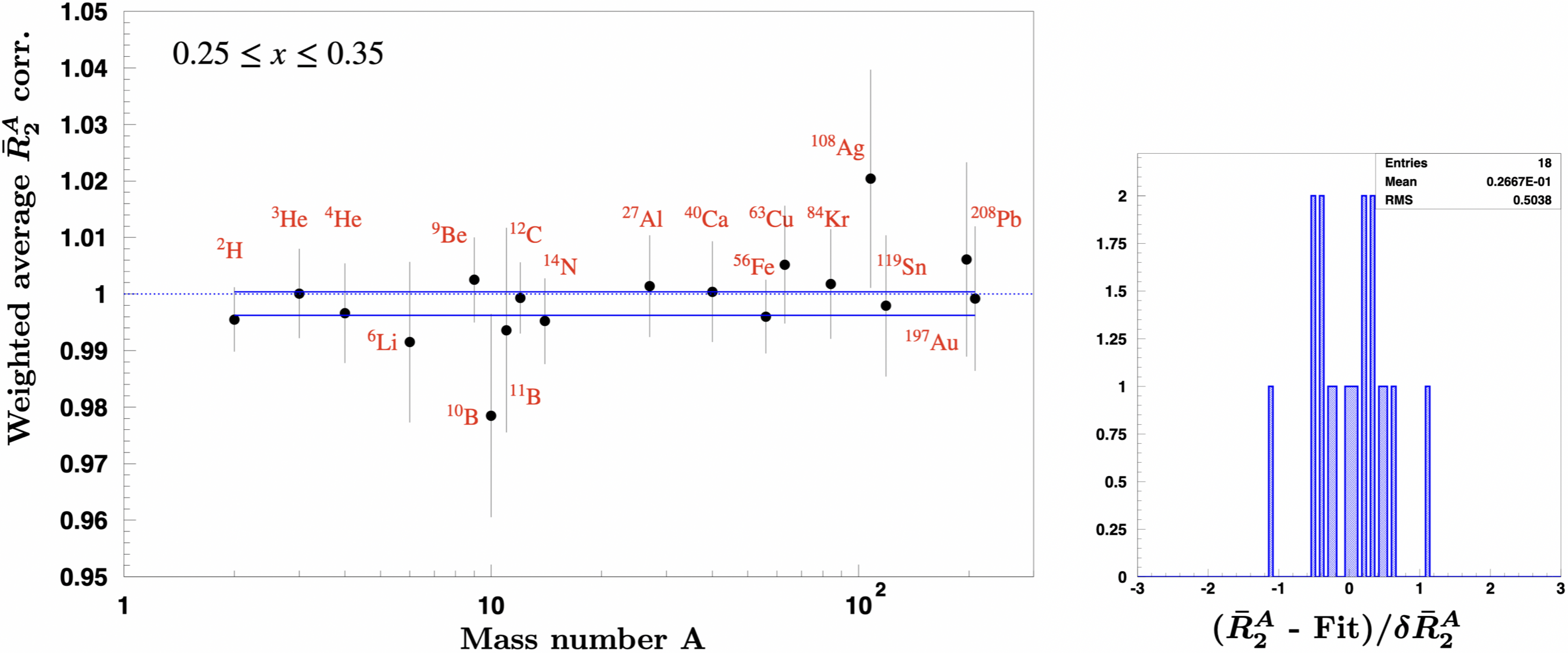}%
\caption{%
Same as Fig.~\ref{fig:emcx03}, but with consistent isoscalarity corrections based on the $F_2^n/F_2^p$ measurements from MARATHON~\cite{JeffersonLabHallATritium:2021usd} and NMC~\cite{NewMuon:1991exl} applied to all $R_2^A$ data points in Table~\ref{tab:avg}. 
}  
\label{fig:emcx03corr}
\end{figure}

\section{Results and discussion} 
\label{sec:discus}

%\vspace*{-4ex}
\begin{figure}[htb]
\centering
\includegraphics[width=0.67\textwidth]{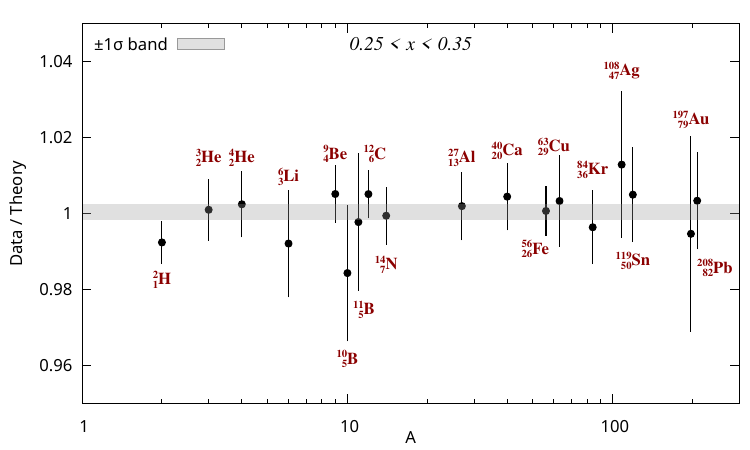}%
\caption{%
Ratio of the averaged $R_2^A$ measurements to the corresponding microscopic model predictions from Refs.~\cite{Kulagin:2004ie,Kulagin:2010gd,Kulagin:2014vsa} for the interval $0.25 \leq x \leq 0.35$. The weighted averages for both experimental data and theoretical calculations are computed according to \eq{eq:avg2}. The shaded area represents the $\pm 1\sigma$ uncertainty band obtained from a constant fit to the full dataset.
}
\label{fig:datth}
\end{figure}

The analysis presented in Sec.~\ref{sec:data} indicates that DIS data demonstrate a remarkable cancellation of nuclear effects when averaged over the region $0.25 \leq x \leq 0.35$. This observation holds across all nuclear targets studied and spans a wide range of four-momentum transfers, from $Q^2 \sim 2 \gevsq$ in JLab measurements to $Q^2 \sim 90 \gevsq$ in CERN experiments.

To understand the cancellation observed, we calculated the predictions of the microscopic model of  Refs.~\cite{Kulagin:2004ie,Kulagin:2010gd,Kulagin:2014vsa} for each data point in Table~\ref{tab:avg}. This model incorporates several key nuclear effects: Fermi motion and binding (FMB) through the smearing with the energy-momentum distribution of bound nucleons, off-shell (OS) modifications to the bound-nucleon structure functions, contributions from meson exchange currents (MEC), and nuclear shadowing (NS) arising from the propagation of the hadronic component of the virtual intermediate boson within the nuclear medium. The theoretical predictions were subject to the same weighted averaging procedure in \eq{eq:avg2} as applied to the experimental data. Figure~\ref{fig:datth} displays the ratios of the averaged $F_2^{A\text{(is)}}/F_2^{{}^2\text{H}}$ data (Table~\ref{tab:avg}) to the corresponding model predictions for various nuclei. Overall, we observe excellent agreement between the model and the data. A fit to the ratios for all nuclei in Fig.~\ref{fig:datth} with a constant yields a value of $1.0004 \pm 0.0022$. In accordance with the discussion in Sec.~\ref{sec:data}, the MARATHON data~\cite{JeffersonLabHallATritium:2024las} are excluded from Fig.~\ref{fig:datth}. Nonetheless, the ratios between the weighted averages of the MARATHON measurements and the model predictions are found to be $0.9996 \pm 0.0073$ for $^3$H/$^2$H and $0.9983 \pm 0.0073$ for $^3$He/$^2$H, demonstrating consistency with the global trend.

\begin{figure}[htb]
\centering
\includegraphics[width=0.5\textwidth]{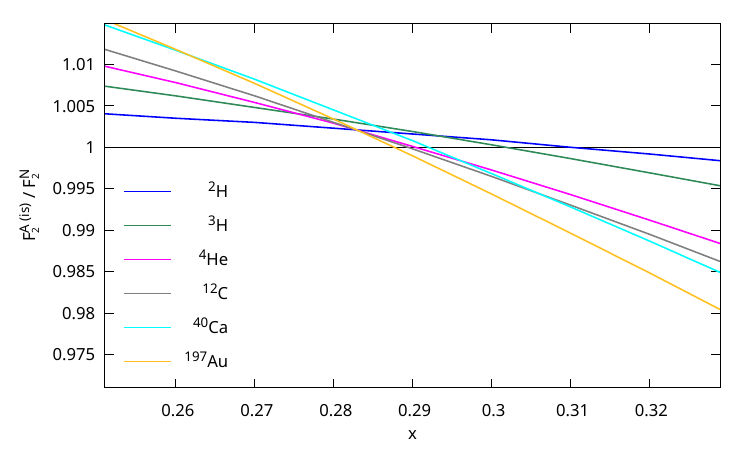}%
\includegraphics[width=0.5\textwidth]{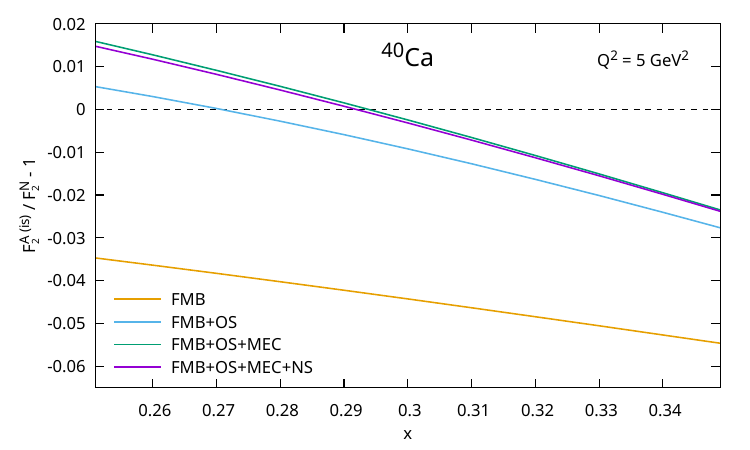}
\caption{%
(Left) Predicted ratios $R_2^{A\text{(is)}}$ for ${}^2$H, ${}^3$H, ${}^4$He, ${}^{12}$C, ${}^{40}$Ca, and ${}^{197}$Au nuclei as a function of $x$ at $Q^2=5\gevsq$. Calculations are based on the microscopic model of Refs.~\cite{Kulagin:2004ie,Kulagin:2010gd,Kulagin:2014vsa}. (Right) Decomposition of the various nuclear contributions to the ratio $R_2^{A\text{(is)}}$ for the ${}^{40}$Ca nucleus: nuclear smearing with the spectral function (FMB); the addition of off-shell corrections (FMB+OS); the inclusion of meson exchange currents (FMB+OS+MEC); and the total result including nuclear shadowing (FMB+OS+MEC+NS).
}
\label{fig:x0:1}
\end{figure}

To investigate the physical origin of the vanishing nuclear corrections near the peak of the valence quark distributions, we calculated the ratios $R_2^{A\text{(is)}}(x,Q^2)$ for various nuclei in the interval $0.25 < x < 0.35$ using the microscopic model from Refs.~\cite{Kulagin:2004ie,Kulagin:2010gd,Kulagin:2014vsa}. Results are displayed in Fig.~\ref{fig:x0:1} (left panel) as a function of $x$ at $Q^2 = 5\gevsq$ for targets ranging from ${}^2$H to ${}^{197}$Au. Notably, the curves for all nuclei intersect in the vicinity of a common point where $R_2^{A\text{(is)}} \approx 1.002$. This deviation from unity is further suppressed for the experimental observable $R_2^A = R_2^{A\text{(is)}} / R_2^{{}^2\text{H(is)}}$, which approaches unity due to the normalization relative to the deuterium. Furthermore, the nearly linear $x$-dependence within this kinematic region results in an enhanced cancellation of nuclear effects when averaged over the range $0.25 \leq x \leq 0.35$. Although the crossover point $x_0$, defined by $R_2^{A\text{(is)}} = 1$, is not identical for all nuclei, it is confined to a remarkably narrow interval around $x_0 = 0.3$.

\begin{figure}[htb]
\centering
\includegraphics[width=0.5\textwidth]{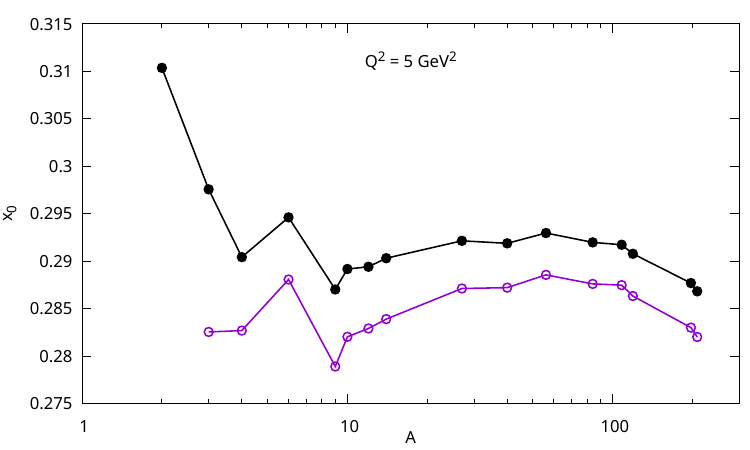}%
\includegraphics[width=0.5\textwidth]{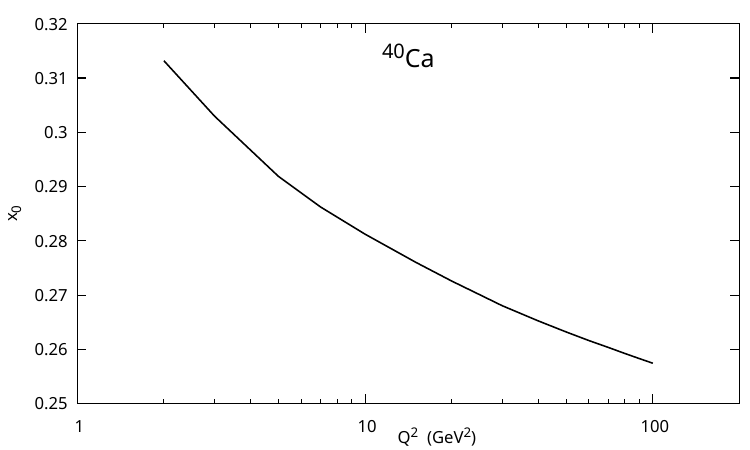}
\caption{%
(Left) The crossover point $x_0$ as a function of the nuclear mass number $A$, computed at $Q^2=5\gevsq$ for the nuclei shown in Fig.~\ref{fig:datth}. The solid symbols correspond to the condition $R_2^{A(\text{is})}(x,Q^2)=1$, while the open symbols represent the solution to $F_2^{A(\text{is})}(x,Q^2)=F_2^{{}^2\text{H}}(x,Q^2)$. 
(Right) The $Q^2$-dependence of the crossover point $x_0$ for the ${}^{40}$Ca nucleus.
}
\label{fig:x0:2}
\end{figure}

It is instructive to evaluate the individual nuclear corrections to $R_2^{A(\text{is})}$ arising from different mechanisms~\cite{Kulagin:2004ie,Kulagin:2010gd,Kulagin:2014vsa}. Figure~\ref{fig:x0:1} (right panel) displays such  contributions for the ${}^{40}$Ca nucleus. In the kinematic region of interest, $0.25 \leq x \leq 0.35$, nuclear DIS is dominated by the incoherent scattering from bound nucleons. The leading nuclear effects include the smearing of the proton and neutron structure functions (SFs) with the nuclear energy-momentum distribution (nuclear spectral function) and the off-shell correction to the bound-nucleon SFs. Notably, the nuclear spectral function employed in this model~\cite{Kulagin:2004ie} accounts for both the mean-field component and the high-momentum tail arising from short-range nucleon-nucleon correlations. We observe a significant cancellation between the nuclear smearing and the off-shell correction in the vicinity of $x=0.3$. The sub-leading contribution from nuclear meson exchange currents therefore also plays a critical role in determining the crossover point $x_0$, as illustrated in Fig.~\ref{fig:x0:1} (right panel). Remarkably, the $A$-dependence of the MEC contribution is found to be similar to that of the nuclear smearing and off-shell effects. This behavior can be understood as the MEC contribution is fundamentally related to nuclear binding through sum rules~\cite{Kulagin:2004ie,Kulagin:1989mu}. As a result, the $A$-dependence of the crossover point $x_0$ is relatively weak, as shown in the left panel of Fig.~\ref{fig:x0:2} (solid symbols). For comparison, we also show the crossover points defined by $F_2^{A(\text{is})}(x,Q^2)=F_2^{{}^2\text{H}}(x,Q^2)$ for the experimentally measurable nuclear ratios relative to deuterium (open symbols). Finally, the $Q^2$-dependence of $x_0$ for ${}^{40}$Ca is illustrated in the right panel of Fig.~\ref{fig:x0:2}.

To further investigate the interplay between the nuclear smearing and the off-shell corrections in the valence region, we consider an approximation for the nuclear convolution framework~\cite{Kulagin:1989mu,Kulagin:1994fz}. Since the typical bound-nucleon energy $\varepsilon$ and momentum $p$ are small compared to the nucleon mass $M$, the nucleon structure function $F_2^N$ within the convolution integral can be expanded in a power series in $\varepsilon/M$ and $\bm p/M$. Smearing this expansion with the nuclear spectral function yields the following expression for isoscalar nuclei:
\begin{equation}\label{eq:dexp}
	R_2^{A(\text{is})} \approx 1 -  e_A x \partial_x \ln F_2^N(x,Q^2) + v_A \delta f(x),
\end{equation}
where $e_A = \langle\varepsilon\rangle/M$ represents the average bound-nucleon energy and $v_A = \langle p^2\rangle/M^2 - 1$ is the average nucleon virtuality, with the nucleon four-momentum defined as $p=(M+\varepsilon,\bm p)$. Note that $\varepsilon$ includes the nucleon separation energy $E>0$ and the recoil energy of the spectator nucleus, which balances the nucleon momentum, $\varepsilon = -E - \bm p^2/(2M_{A{-}1})$.
At the leading order in $1/M$ the nucleon virtuality can be expressed as $v_A = 2(\langle \varepsilon \rangle - \langle T \rangle)/M$ in terms of the average nucleon energy $\la\ceps\ra$ and kinetic energy $\langle T\rangle = \langle\bm p^2\rangle/(2M)$. The averaging is performed using the nuclear spectral function as detailed in Ref.~\cite{Kulagin:2004ie}. In \eq{eq:dexp}, we neglect the second-derivative term, $\langle T\rangle x^2\partial_x^2F_2$, as its contribution is small in the kinematic region under consideration.
The function $\delta f = M^2 \partial_{p^2} \ln F_2^N(x,Q^2,p^2)$ denotes the derivative with respect to the invariant mass squared $p^2$ of the off-shell nucleon near the mass shell $p^2=M^2$~\cite{Kulagin:2004ie,Kulagin:2010gd}. This function describes the relative modification of the nucleon structure functions in the off-shell region and is, by definition, a universal quantity independent of the specific nucleus. Following this framework, off-shell modifications in nuclear DIS have been studied extensively~\cite{Kulagin:2004ie,Kulagin:1994fz,Kulagin:2010gd,Kulagin:2014vsa,Alekhin:2017fpf,Alekhin:2022tip,Alekhin:2022uwc}. A $Q^2$-independent approximation, $\delta f(x)$, has been shown to provide an excellent description of nuclear DIS data~\cite{Kulagin:2004ie,Kulagin:2010gd,Alekhin:2017fpf,JeffersonLabHallATritium:2024las}, Drell-Yan lepton-pair production~\cite{Kulagin:2014vsa}, and $W^\pm/Z$ boson production in $p+\text{Pb}$ collisions at the LHC~\cite{Ru:2016wfx}. Furthermore, a recent global QCD analysis~\cite{Alekhin:2022uwc} found the proton-neutron asymmetry, $\delta f^p - \delta f^n$, to be consistent with zero within uncertainties. We therefore employ a single, universal off-shell function $\delta f(x)$ for both protons and neutrons, as determined from the global study of nuclear SFs in Ref.~\cite{Kulagin:2004ie}.

From \eq{eq:dexp} the binding and off-shell corrections cancel out if:
\begin{equation}\label{eq:x}
x \partial_x \ln F_2^N(x,Q^2) =  a \delta f(x),
\end{equation}
where $a = v_A / e_A$. If the parameter $a$ were independent of the mass number $A$, \eq{eq:x} would yield a universal solution $x_0$ for all nuclei. This condition is satisfied if $e_A$ and $v_A$ exhibit a similar $A$-dependence, which then cancels out in their ratio. In this context, we note that for a nucleus described by a Hamiltonian with two-body interactions, $\la\ceps\ra$ and $\la T\ra$ are related by the Koltun sum rule~\cite{Koltun:1972kh}:
\begin{equation}\label{eq:ksr}
\la\ceps\ra + \la T\ra = 2\ceps_0,
\end{equation}
where $\ceps_0 < 0$ is the nuclear ground-state energy per nucleon. Applying \eq{eq:ksr}, the ratio $a$ can be expressed as:
\begin{equation}\label{eq:a}
a=4\left(1 - \ceps_0/\la\ceps\ra\right).
\end{equation}
Since the ratio $\ceps_0 / \la\ceps\ra$ is small for all nuclei, the quantity $a$ exhibits only a weak $A$-dependence. For the deuteron, using $\ceps_0 = -1.11 \mev$ and the deuteron wave function with the AV18 nucleon-nucleon potential~\cite{Wiringa:1994wb,Veerasamy:2011ak}, we obtain $\la\ceps\ra = -11.9 \mev$ and $a = 3.63$. For complex nuclei, we utilize a model spectral function~\cite{Kulagin:2000yw,Kulagin:2004ie} to compute $\la\ceps\ra$, finding that $a$ spans a narrow range from approximately $3.5$ to $3.2$ for nuclei ranging from $^3$H ($^3$He) to ${}^{208}$Pb.

Solving \eq{eq:x} numerically at $Q^2 = 5 \gevsq$, we obtain $0.26 < x_0 < 0.31$ for $^2$H, $^3$H, $^4$He, $^{12}$C, $^{40}$Ca, and $^{197}$Au targets, with the maximum value of $x_0$ occurring for deuterium and the minimum for gold. The MEC correction significantly shifts $x_0$ toward higher values for heavy targets (as illustrated for $^{40}$Ca in the right panel of Fig.~\ref{fig:x0:1}), whereas its impact on the deuteron is negligible. Consequently, the inclusion of MEC reduces the overall spread of $x_0$ across different nuclei, resulting in the behavior displayed in the left panel of Fig.~\ref{fig:x0:1}.

The crossover point $x_0$ evolves toward lower values as $Q^2$ increases, a trend driven primarily by the $Q^2$-evolution of the nucleon structure function $F_2^N$. Notably, the $A$-dependence of $x_0$ is significantly weaker than its $Q^2$-dependence, as illustrated in the right panel of Fig.~\ref{fig:x0:2}. However, existing experimental data on the $Q^2$-dependence remain sparse, both in kinematic reach and target variety. New high-precision measurements spanning a broad scale of $Q^2 \sim 1\text{--}100\gevsq$ across multiple nuclei within a single experiment are therefore highly desirable. To unambiguously resolve these predicted shifts, the critical normalization uncertainty is required to be below 
1\%. Future programs at the Electron-Ion Collider~\cite{AbdulKhalek:2021gbh} and long-baseline neutrino facilities~\cite{DUNE:2020ypp,Petti:2022bzt} possess the potential sensitivity needed for such measurements.

\acknowledgments
We thank S. I. Alekhin and G. G. Petratos for useful discussions. This work is supported by Grant No. DE-SC0026395 from the Department of Energy, USA.

%\newpage
\appendix
\section{Non-isoscalarity corrections and $F_2^n/F_2^p$}
\label{sec:F2n-F2p}

Comparing the $R_2$ ratios for non-isoscalar nuclei ($Z \neq N$) requires a correction for the neutron or proton excess, as outlined in Sec.~\ref{sec:data}. This non-isoscalarity correction depends explicitly on the ratio of the free nucleon structure functions, $R_{np} = F_2^n/F_2^p$. Because the $R_{np}$ models adopted by different experimental collaborations are not always mutually consistent, they introduce an additional source of systematic uncertainty. In the analysis of Sec.~\ref{sec:data}, we evaluated the impact of such a model dependence by replacing the original non-isoscalarity corrections applied by the experiments with a consistent correction derived from a parameterization of $R_{np}$ based on MARATHON~\cite{JeffersonLabHallATritium:2021usd} and NMC~\cite{NewMuon:1991exl} measurements. 

\begin{figure}[htb!]
\centering
\includegraphics[width=0.8\textwidth]{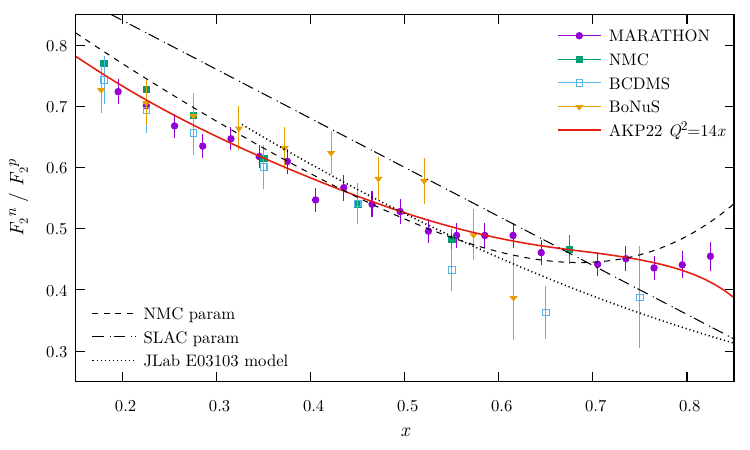}%
\caption{%
Ratio $R_{np}= F_2^n / F_2^p$ measured by the MARATHON experiment~\cite{JeffersonLabHallATritium:2021usd} compared with data from NMC~\cite{NewMuon:1996uwk}, BCDMS~\cite{BCDMS:1989gtb}, and BoNuS~\cite{CLAS:2014jvt} ($W \geq 1.8\gev$), all evolved to the MARATHON kinematic scale of $Q^2 = 14x \gevsq$. The solid red curve displays the calculation using the proton and neutron structure functions from the AKP22 global QCD fit~\cite{Alekhin:2022uwc}. For comparison, the phenomenological parameterizations from NMC~\cite{NewMuon:1991exl} and SLAC~\cite{Gomez:1993ri} are shown alongside the JLab E03-103 model prediction~\cite{Arrington:2021vuu}, each evolved to the same scale.
}
\label{fig:Rnp-comp}
\end{figure}

Figure~\ref{fig:Rnp-comp} displays a comparison of the 
recent $R_{np}$ measurement by MARATHON~\cite{JeffersonLabHallATritium:2021usd} with the corresponding determinations by NMC~\cite{NewMuon:1996uwk}, BCDMS~\cite{BCDMS:1989gtb}, and BoNuS~\cite{CLAS:2014jvt}. These measurements were based on different experimental techniques and targets. The MARATHON experiment extracted $R_{np}$ exploiting the partial cancellation of nuclear effects in the ratio of structure functions of $A=3$ mirror nuclei, $F_2^{{}^3\text{H}}/F_2^{{}^3\text{He}}$. The resulting uncertainties introduced by the nuclear corrections on the $R_{np}$ determination are substantially smaller than the corresponding experimental uncertainties~\cite{Afnan:2000uh, Afnan:2003vh}. The NMC and BCDMS experiments obtained $R_{np}$ from measurements of the $F_2^{{}^2\text{H}}/F_2^p$ ratio in DIS using high-energy muon beams at CERN. The BoNuS experiment determined $R_{np}$ from the ratio $F_2^n/F_2^{{}^2\text{H}}$ measured with an electron beam at lower CEBAF energies. The $F_2^n$ was obtained from DIS on ${}^2\text{H}$ by tagging spectator protons to select barely off-shell neutrons. 

Because the datasets considered are recorded at different $Q^2$ scales for a given value of $x$, a direct comparison requires a consistent scale transformation. We evolved the NMC, BCDMS, and BoNuS data to the reference scale of the most recent MARATHON measurement, $Q^2_M = 14x \gevsq$, by applying the correction factor $R_{np}(x,Q_M^2)/R_{np}(x, Q^2)$ 
to each data point $(x, Q^2)$. As a result, we observe excellent consistency among all measurements within the valence region of interest, $0.2 < x < 0.4$ (Fig.~\ref{fig:Rnp-comp}). The prediction from the AKP22 global QCD fit~\cite{Alekhin:2022uwc} shows good agreement with both the MARATHON and NMC measurements over the full kinematic range, thereby providing a robust parametrization of the data. For completeness, Fig.~\ref{fig:Rnp-comp} also displays the phenomenological parametrization obtained from the NMC fit to NMC, BCDMS, and SLAC $R_{np}$ data~\cite{NewMuon:1991exl}, the legacy parametrization used in the SLAC E139 analysis~\cite{Gomez:1993ri}, and the JLab E03-103 model~\cite{Arrington:2021vuu}, all evolved to $Q^2 = Q_M^2$.

Despite the significant discrepancies among the $R_{np}$ parameterizations and models used by the various  collaborations, replacing individual non-isoscalarity corrections with a consistent one based on a parameterization of the MARATHON~\cite{JeffersonLabHallATritium:2021usd} and NMC~\cite{NewMuon:1991exl} measurements has a marginal impact on the results of our analysis in Sec.~\ref{sec:data}. The shift in the fitted constant $R_0$ when swapping from the original published data to the uniformly corrected ones is notably smaller than the corresponding fit uncertainty, as shown in Table~\ref{tab:fitsA}. Furthermore, both fits are in excellent agreement with the fit restricted solely to isoscalar nuclei. We can therefore conclude that the specific treatment of the non-isoscalarity corrections does not introduce any significant systematic bias into the results of the present work.

\section{Normalization of experimental data}
\label{sec:datanorm}

Most measurements listed in Table~\ref{tab:avg} are characterized by relatively large normalization uncertainties, which typically dominate over the statistical and point-to-point systematic errors. Factors contributing to this normalization uncertainty include beam luminosity calibrations, estimations of the cryogenic target density, radiative corrections, and detector acceptance corrections. Conversely, a few experiments were specifically designed to achieve a precise determination of the absolute normalization. The NMC experiment at CERN, for example, achieved an overall normalization uncertainty of 0.4\% on its $\sigma^A/\sigma^{{}^2\text{H}}$ measurements~\cite{Amaudruz:1995tq,NewMuon:1996yuf}. The use of a muon beam significantly reduced both target heat dissipation and the corresponding uncertainty on cryogenic target densities, while minimizing the impact of radiative corrections relative to electron-beam experiments. Furthermore, NMC implemented rigorous luminosity monitoring and physically rotated the targets to eliminate geometric acceptance and efficiency corrections. The SLAC E140 experiment, although utilizing an electron beam, was specifically designed to provide a reference normalization for all SLAC DIS measurements, achieving a 1.1\% normalization uncertainty on the $\sigma^A/\sigma^{{}^2\text{H}}$ ratios~\cite{Dasu:1993vk}. More recently, the MARATHON experiment at JLab attained a 0.55\% normalization uncertainty on  $\sigma^{{}^2\text{H}}/\sigma^\text{H}$~\cite{JeffersonLabHallATritium:2021usd}, which served to calibrate its measurements on other nuclear targets. This experimental setup relied on high-pressure gaseous targets cycled frequently for each kinematic setting to minimize instrumental drifts and mitigate acceptance and efficiency variations.

Cross-normalizing new datasets to either higher-precision reference measurements or world-average cross sections is an established practice across various experimental sub-fields (see, e.g., Ref.~\cite{ParticleDataGroup:2024cfk}). This procedure is particularly beneficial when the intrinsic normalization uncertainty of a dataset exceeds its combined statistical and point-to-point systematic errors. A prerequisite for this approach is that the reference and target measurements share an overlapping kinematic domain. Because the target data are scaled to a fixed external baseline, the resulting normalization uncertainty of the cross-normalized dataset becomes identically that of the reference input.

Following this strategy, the HERMES experiment at DESY utilized a cross-normalization to NMC data to address its normalization uncertainty originating from the calibration of the beam luminosity monitor~\cite{Garutti:2003}. Because the beam luminosity is common to all targets, a uniform scaling factor of 0.9\% was determined for all $\sigma^A/\sigma^{{}^2\text{H}}$ measurements by comparing HERMES and NMC data within their overlapping $x$ region. This 0.9\% adjustment is statistically consistent with the intrinsic 1.4\% HERMES normalization uncertainty, which was consequently replaced by the tighter 0.4\% normalization uncertainty of the NMC reference baseline.

Similarly, the MARATHON experiment at JLab employed an internal cross-normalization procedure~\cite{JeffersonLabHallATritium:2021usd,JeffersonLabHallATritium:2024las} to scale the $\sigma^{A^\prime}/\sigma^{A}$ ratios to its high-precision $\sigma^{{}^2\text{H}}/\sigma^p$ measurement, as they shared identical kinematics and detector configurations. This procedure requires consistency among $F_2^n/F_2^p$ values extracted from different sets of nuclear DIS ratios near $x=0.3$. Crucially, this cross-normalization is entirely data-driven and independent of model predictions. The sole underlying physical assumption is that nuclear corrections near $x=0.3$ are significantly smaller than the corresponding experimental uncertainties~\cite{Afnan:2000uh, Afnan:2003vh}. We have verified the validity of this assumption in the main text.
Since nuclear modifications are demonstrated to be on the order of a few per-mille within the range $0.25 \leq x \leq 0.35$, their impact on MARATHON's internal cross-normalization is negligible.

In contrast, the $\sigma^A/\sigma^{{}^2\text{H}}$ measurements from the JLab E03-103 experiment~\cite{Arrington:2021vuu,Seely:2009gt} carry a comparatively large normalization uncertainty: 1.84\% for ${}^3$He and ranging from 1.60\% to 2.02\% for heavier nuclear targets. Dominant contributions to these scale uncertainties stem from the absolute target thicknesses, cryogenic target density variations, and  radiative or background corrections. Once these normalization uncertainties are properly accounted for, the JLab E03-103 measurements are found to be statistically compatible with global data~\cite{Kulagin:2010gd}.

%%% REFERENCES
\bibliography{main}

\end{document}